\documentclass[12pt,twocolumn]{article}

\usepackage{amsmath}
\usepackage{amssymb}
\usepackage{graphicx}
\usepackage{xcolor}
\usepackage{siunitx}
\usepackage{enumerate}
\usepackage[margin=0.75in]{geometry}
\usepackage{url}
\usepackage{natbib}

\title{Venting and Outgassing Simulations of Pressurized Lunar Modules: Contamination of the Lunar Environment}

\author{S. Boccelli$^1$, W.M. Farrell$^{2,3}$, P. Saxena$^1$ \& O.J. Tucker$^{1,4}$ \\
{\footnotesize $^1$ NASA Goddard Space Flight Center, 8800 Greenbelt Rd, Greenbelt, MD 20771, USA}\\
{\footnotesize $^2$ Space Science Institute, 4765 Walnut Street, Suite B, Boulder, CO 80301, USA}\\
{\footnotesize $^3$ DeepSpace Technologies Inc., 8865 Stanford Boulevard, Suite 183, Columbia, MD 21045, USA}\\
{\footnotesize $^4$ Department of Atmospheric and Planetary Sciences, Hampton University, Hampton, VA 23668, USA}}

\date{ }

\begin{document}

% \linenumbers

\twocolumn[
  \begin{@twocolumnfalse}

  \maketitle

  \begin{abstract}
  One objective of Artemis science is to determine the impact human activities have on the lunar environment, which might compromise science objectives and measurements. 
  We perform a preliminary analysis of the contamination associated with airlock venting and outgassing from a prototype lunar-module geometry intended to host astronauts 
  on the lunar surface. 
  The air flow generated by the depressurization of the airlock, expanding in the lunar exosphere,
  is studied using the Direct Simulation Monte Carlo (DSMC) method for two different venting configurations and the 
  particle flux on the surface is computed as a function of the distance from the the module.
  Outgassing from the main body of the module---assumed to be covered with a Multi-Layer Insulation (MLI) blanketing---and from the solar panels is then
  analyzed using a view-factor method, employing outgassing rates from the literature.
  Our results give preliminary indications of the distance at which contamination levels fall below the values characteristic of native species in the lunar atmosphere.
  Scientific measurements targeting ${}^{40}\mathrm{Ar}$ should be carried farther than $30$--$100$ meters from the module, while the detection of lower-abundance species
  such as polar-crater water might require to travel up to and beyond $3~\si{km}$ from the module.
  \end{abstract}

  \end{@twocolumnfalse}
]

% =====================
% =====================
% =====================

\section{Introduction}

As lunar exploration efforts increase worldwide 
\citep{schonfeld2023summary,smith2020artemis,li2019china,mathavaraj2024chandrayaan,mitrofanov2021luna}, 
the need to quantify the effect of antropogenic activities
on the lunar exosphere and regolith environment becomes prioritary \citep{killen2024moon}.
In-situ resource utilization (ISRU) activities \citep{zhang2023overview},
deposition of chemicals from propulsion-system plumes \citep{farrell_lingering_2022} and propellant venting procedures are contamination sources
that might interfere with scientific experiments \citep{A3definitionteamreport,saxena_situ_2023} and thus require careful 
evaluation.
As observed during the Apollo missions \citep{stern1999lunar},
water outgassing from spacecraft, rovers and mobile surface habitats \citep{kessler2022artemis} 
as well as astronaut suits during Extra Vehicular Activities (EVAs)
might locally exceed the exospheric volatile content \citep{killen_temporary_2024}, 
preventing the succesful execution of mass-spectrometric and chemical-analyzer measurements \citep{cohen2024peregrine,sultana2024alaska}.
Human activity is also a source of biological contamination.
Airlock depressurization processes and carbon-dioxide removal sub-systems embedded in the astronaut Portable Life Support System (PLSS) \citep{chullen2018swing} 
are likely to expel traces of organics and bacteria on the surface \citep{saxena_situ_2023}, interfering with astrobiological investigations \citep{weiss2024operational}.

\begin{figure*}[htpb]
  \centering
  \includegraphics[width=\textwidth]{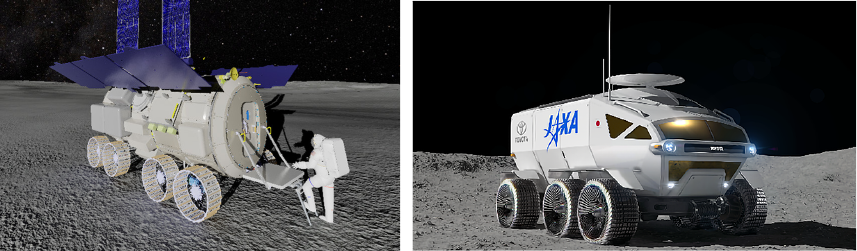}
  \caption{Artist's rendering of the Multi-Purpose Habitation lunar module (MPH, Left) and of the Lunar Cruiser pressurized rover (Right). 
           Credit: Thales Alenia Space / Agenzia Spaziale Italiana (ASI) \citep{MPHthales}, and JAXA \citep{LunarCruiserJAXA}.}
  \label{fig:MPH-artist}
\end{figure*}

Pressurized mobile structures, serving as habitats for astronauts, are presently being developed in support of NASA's Artemis mission.
These include the Multi-Purpose Habitation (MPH) lunar module \citep{iliano2024international,overview2025parodi}, developed by the Italian Space Agency (ASI) and its
prime industrial contractor Thales Alenia Space, as well as the Lunar Cruiser pressurized rover \citep{yamazaki2024overview,yamaguchiconceptual}
by the Japan Aerospace Exploration Agency (JAXA) and Toyota. 
These modules are shown in Fig.~\ref{fig:MPH-artist}. 
Their main features include a pressurized habitation section, an airlock, and large solar panels---retracted and not visible in the 
Lunar Cruiser illustration of Fig.~\ref{fig:MPH-artist}.

In this work we build a prototype astronaut-module geometry and perform an external gas dynamic analysis of the airlock venting process
in order to characterize the possible spacial extent of surface contamination. 
Also, we attempt an estimation of the contamination caused by outgassing of the solar panels and of the Multi-Layer Insulation (MLI) blanketing, possibly applied on the module's body.
For this analysis we neglect important parameters such as the structure temperature, determined by the thermal design, 
the specific materials employed, the details on the MLI application, time on the lunar day and the exact geometry of the module.
Such parameters are presently not available as the design and the definition of the ConOps are ongoing. 
We attempt nonetheless an order-of-magnitude estimation based on outgassing values typical of existing spacecraft and landers \citep{boccelli2025DSMCanalysis}.

This manuscript is organized as follows.
In Section~\ref{sec:geometry} we build a simplified geometry of the lunar module, inspired on available images of the future MPH.
Section~\ref{sec:depressurization} analyzes the depressurization process using a simple zero-dimensional model.
Certain pressure conditions of interest are identified from the transitory and are analyzed in detail in 
Section~\ref{sec:numerical-simul-venting}, where we simulate the three-dimensional expansion of the vented gas around the module and plot maps illustrating the 
molecular flux to the ground.
In Section~\ref{sec:numerical-simul-outgassing} we perform an outgassing simulation up to 120 meters from the module and 
in Section~\ref{sec:lunar-environment} we compare these results with the natural abundance of species to be expected on the Moon, further extending the simulated domain to
a distance of 5 km. 
The conclusions are drawn in Section~\ref{sec:conclusions}.

%%%%%%%%%%%%%%%%%%%%%%%%%%%%
%%%%%%%%%%%%%%%%%%%%%%%%%%%%
%%%%%%%%%%%%%%%%%%%%%%%%%%%%

\section{A simplified lunar-module geometry}\label{sec:geometry}

\begin{figure*}[htpb]
  \centering
  \includegraphics[width=\textwidth]{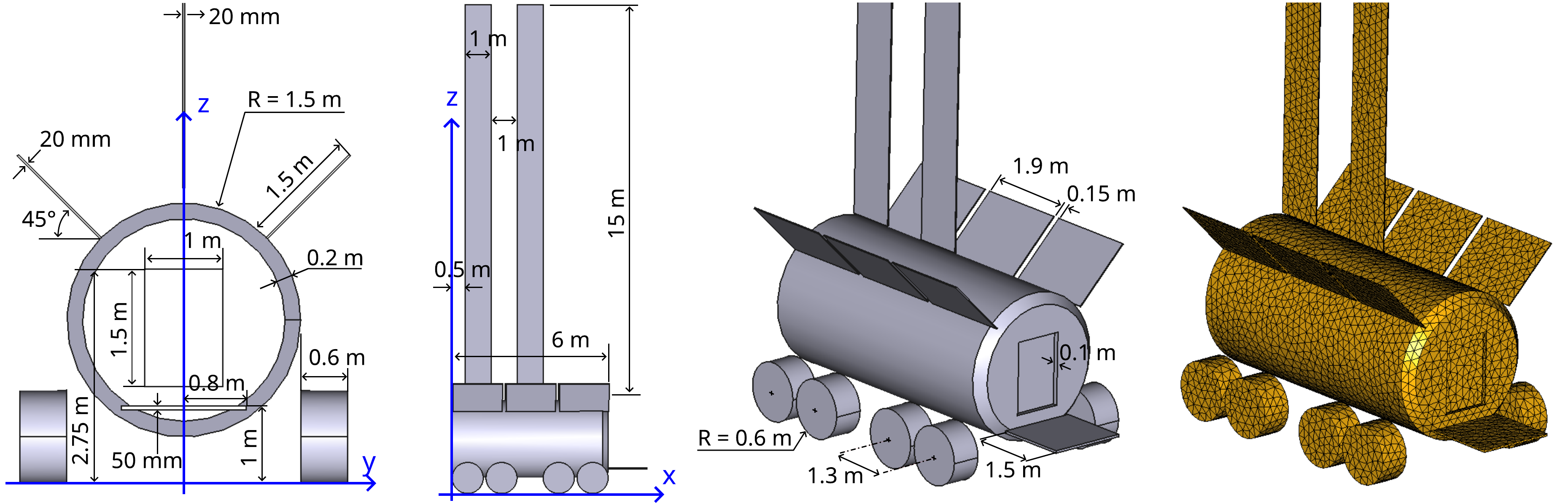}
  \caption{Simplified geometry employed in our simulations.}
  \label{fig:MPH-geometry}
\end{figure*}

The specific geometries of the Lunar Cruiser and of MPH are presently being defined. 
For the Lunar Cruiser the whole cabin vents in the lunar exosphere in preparation for EVAs \citep{yamazaki2024overview}.
In case of MPH, current designs include a pressurized habitation module section and an airlock.
Depending on energy considerations, the depressurization of the airlock might be performed by partly recycling the gas 
using compressors, or might be entirely vented in the exosphere.
In order to simulate this venting process, we have built a simplified geometry shown in Fig.~\ref{fig:MPH-geometry} and loosely based on MPH's 
published images.
In this geometry we take the module to have a diameter $D = 3~\si{m}$ and a total length $L = 6~\si{m}$, 
divided in an habitation section ($L_h = 4~\si{m}$) and an airlock $L_a = 2~\si{m}$.
Correspondingly, the volume of the airlock section is here approximated as $V_a = 14.137~\si{m^2}$.
Our geometry includes two $15~\si{m}$--long solar panels and two radiators angled at $45^\circ$ with respect to the vertical plane.
The solar panels and the radiators are here approximated with a thickness of $2~\si{cm}$ but this specific quantity is irrelevant
to the simulations of this work.
Figure~\ref{fig:MPH-geometry}-Right shows one of the surface grids employed here to compute the surface particle fluxes. 

\begin{figure}[htpb]
  \centering
  \includegraphics[width=\columnwidth]{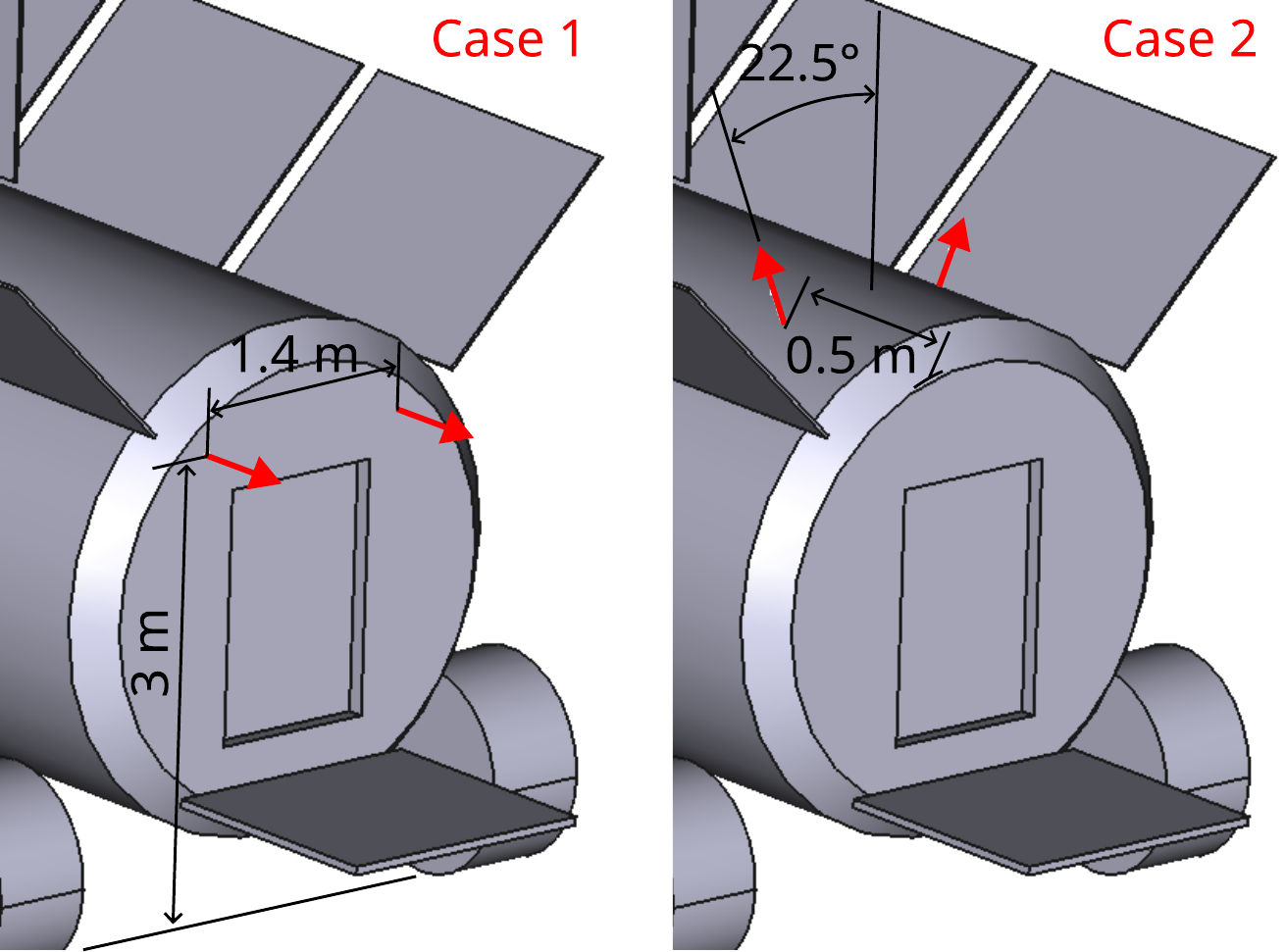}
  \caption{Position of the airlock depressurization vents for the simulations of Case 1 (Left) and Case 2 (Right).}
  \label{fig:position-venting-ports}
\end{figure}

The position and orientation of the airlock venting valves is still undefined.
For existing and proposed airlock designs the reader is referred to \citet{williams2004international} and \citet{vrankar2023airlock}.
We simulate here two venting ports located symmetrically with respect to the $(x,z)$ plane.
Two different configurations are analyzed, Case 1 and Case 2, as shown in Fig.~\ref{fig:position-venting-ports}.
In Case 1 the vents are located above the exit door and are directed horizontally.
Instead, in Case 2 the vents are directed vertically at an angle of $22.5^\circ$ 
(that is intermediate between the vertical solar panels and the radiators, angled at $45^\circ$) and 
are located $0.5~\si{m}$ into the airlock section.
The Case 1 and Case 2 configurations are simulated separately.
Each vent is considered to be a circular aperture of radius $R = 7.5~\si{mm}$ as defined in Section~\ref{sec:depressurization}.

%%%%%%%%%%%%%%%%%%%%%%%%%%%%
%%%%%%%%%%%%%%%%%%%%%%%%%%%%
%%%%%%%%%%%%%%%%%%%%%%%%%%%%

\section{Depressurization process: mass flow rates and vent size}\label{sec:depressurization}

In this section we estimate the outflow rates during the depressurization process and the associated vent area.
These parameters are necessary to run the three-dimensional simulations of Section~\ref{sec:numerical-simul-venting}.
Since MPH and the Lunar Cruiser roughly have a comparable size, our analysis can be somewhat transfered among the two different modules.

Following \citet{overview2025parodi}, we consider an initial gas pressure of $P_{a0} = 56.5~\si{kPa}$, 
representative of the design operating pressure of the habitat section.
We assume the gas is an oxygen-nitrogen mixture with an oxygen molar fraction $X_\mathrm{O_2} = 0.3$.
This satisfies the requirement that $X_\mathrm{O_2} \le 0.37$ \citep{overview2025parodi}.
The resulting mixture has an individual gas constant
\begin{equation}
  R_i = \frac{k_B}{X_\mathrm{N_2} m_\mathrm{N_2} + X_\mathrm{O_2} m_\mathrm{O_2}} = 284.75~\si{\frac{J}{kg\, K}}\, ,
\end{equation}

\noindent where $k_B$ is the Boltzmann constant.
The equivalent (weighted) molecular mass of the mixture is $m_\mathrm{mix} = 4.85\times10^{-26}~\si{kg}$ 
and the ratio of specific heats is $\gamma = 1.4$ (only the translational and 
rotational modes are activated \citep{ferziger1972mathematical}).
The initial density in the airlock is thus $\rho_{a0} = 0.6614~\si{kg/m^3}$.

The evolution of $\rho_a$ during the depressurization can be estimated from a simple zero-dimensional mass balance,
\begin{equation}\label{eq:zero-dimensional-model}
  \frac{\mathrm{d} \rho_a}{\mathrm{d} t} = - \frac{\dot{m_v}}{V_a} \, ,
\end{equation}

\noindent where $V_a$ is the volume of the airlock and $\dot{m}_v = 2 \rho_v u_v A_v$ is the mass flux outflowing from two circular vents, each with an area $A_v = \pi R_v^2$,
$R_v$ being an effective radius for each individual valve.
The gas properties at the vent are estimated here using the adiabatic expansion formulas \citep{liepmann2001elements},
\begin{equation} 
  \begin{cases}
  T_v    = T_a \left( 1 + \frac{\gamma - 1}{2} M_v^2 \right)^{-1} \, , \\
  \rho_v = \rho_a \left( 1 + \frac{\gamma - 1}{2} M_v^2 \right)^{-1/(\gamma - 1)} \, , \\
  u_v = M_v \sqrt{\gamma R_i T_v} \, ,
  \end{cases}
\end{equation}

\noindent where $M_v$ is the Mach number reached at the vent during the expansion and $T_a$ and $P_a$ 
are the temperature and pressure in the airlock.
As the external pressure is negligible, near the vent the gas reaches a sonic speed (Mach number $\mathrm{M}_v=1$).
This condition completely defines the outflow velocity, $u_v$, and lets us compute the mass flow rate.
Notice that the gas at the vent reaches $\mathrm{M}_v=1$ both during the initial phases of the expansion---when the pressure is high---and also at the end of the depressurization phase,
when the gas is rarefied (see for instance \citet{sharipov2004numerical}).

In order to further simplify the analysis we assume that the temperature in the airlock remains constant during the depressurization 
and is equal to $T_a = 300~\si{K}$. 
This simplifying assumption is expected to affect the specific shape of the pressure curve.
However, this does not constitute an issue for the present study, since we are merely concerned with estimating 
reasonable values for the mass flow rates, and we are not targeting an accurate design of the depressurization process.
Ultimately, since the internal temperature is held constant, the temperature at the vent also remains constant and equal to 
$T_v = 250~\si{K}$ throughout the expansion, and the corresponding velocity at $\mathrm{M}_v=1$ is $u_v = 315.69~\si{m/s}$.

\begin{figure}[htpb]
  \centering
  \includegraphics[width=\columnwidth]{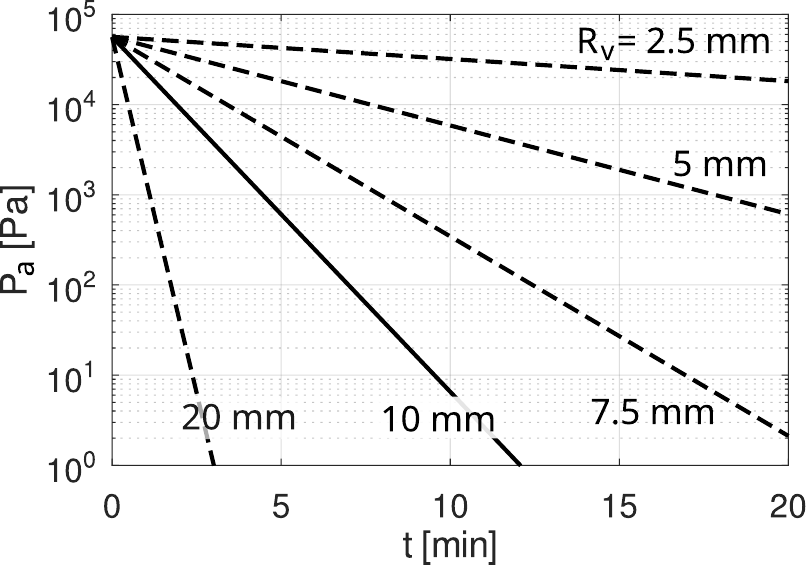}
  \caption{Airlock depressurization curves during venting from two valves each with an equivalent radius $R_v$.
Only the case $R_v = 10~\si{mm}$ is selected in this work while the other curves (dashed lines) are not considered further.}
  \label{fig:discharge-pressure-time}
\end{figure}

\begin{figure}[htpb]
  \centering
  \includegraphics[width=\columnwidth]{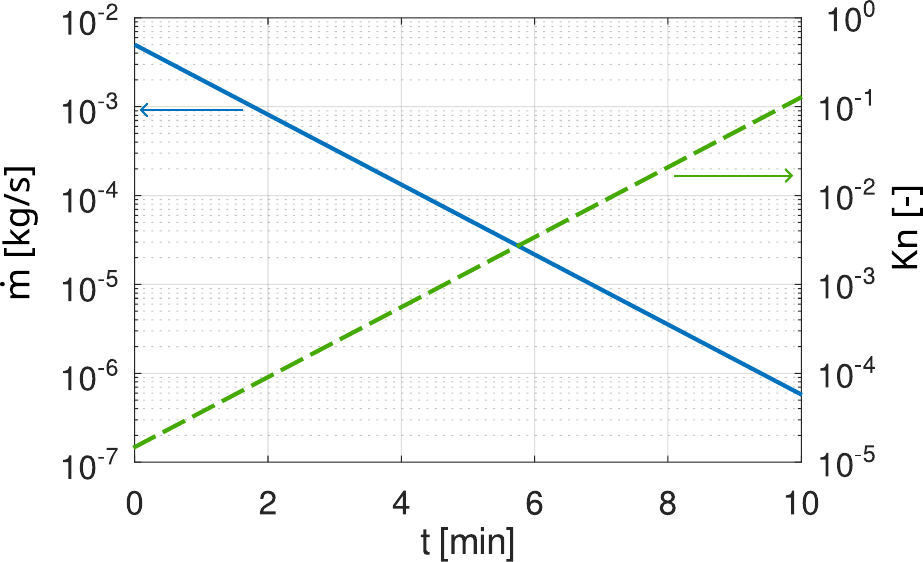}
  \caption{Evolution of the flow through a single vent of radius $R_v = 10~\si{mm}$.
           Solid blue line: mass flow rate through an individual vent. Dashed green line: Knudsen number at the vent location.}
  \label{fig:mdot_vent_time}
\end{figure}

Eq.~\eqref{eq:zero-dimensional-model} is easily solved analytically or can be integrated numerically.
The depressurization curves obtained for different vent radii are shown in Fig.~\ref{fig:discharge-pressure-time}.
Ultimately, depressurization curves will be designed as to meet the required dwell times for astronauts, 
based on health and safety requirements \citep{abercromby2013fifteen,gernhardt2007biomedical}.
Here, we consider only the case of a vent radius $R_v = 10~\si{mm}$, which evacuates the airlock in about ten minutes.
Other cases are analogous and this choice does not compromise the validity of the results.
At $t = 10~\si{min}$, the residual pressure in the airlock is approximately $10~\si{Pa}$, which permits to effortlessly operate the outer door.
The time-evolution of the mass flow rate passing through a vent with a radius $R_v = 10~\si{mm}$ is shown in Fig.~\ref{fig:mdot_vent_time}, while the 
total flow leaving the airlock is twice as much.

During the airlock depressurization, the density drops and the level of rarefaction of the flow through the vent increases.
An order-of-magnitude indication of the importance of rarefied-gas effects can be obtained by analyzing the Knudsen number,
$\mathrm{Kn}$, expressing the ratio of the collision mean free path, $\lambda$, to the vent radius, $R_v$:
\begin{equation}
  \mathrm{Kn} = \frac{\lambda}{R_v} = \frac{1}{R_v} \frac{k_B T_v}{\sqrt{2} \, \sigma P_v} \, ,
\end{equation}

\noindent where $\sigma$ is the elastic-collision cross-section, which can be estimated using the variable-hard-sphere parameters reported in
\citet{bird1994molecular} for air. 
At $T=300~\si{K}$ one has $\sigma \approx 5.5\times10^{-19}~\si{m^2}$.
The 30\%--70\% oxygen-nitrogen mixture considered here would have a similar value and a small discrepancy would not compromise our Knudsen-number analysis.
The evolution of $\mathrm{Kn}$ at the vent is also shown in Fig.~\ref{fig:mdot_vent_time}.

At the beginning of the depressurization, the density is sufficiently large that the collisional mean free path is small as compared to the 
vent radius.
In this condition, the Knudsen number is approximately $\mathrm{Kn}\approx10^{-5}$, indicating that the flow is in the continuum regime
and can be simulated using continuum fluid dynamic models such as the Euler or the Navier-Stokes equations 
\citep{lofthouse2007effects}.
However, as the pressure decreases, the mean free path eventually becomes comparable with the vent size.
At $t = 10~\si{min}$ the Knudsen number is $\mathrm{Kn} \approx 0.1$, and the flow through the vent should be studied
employing a rarefaction-capable model such as the Direct Simulation Monte Carlo (DSMC) method \citep{bird1994molecular}.
These preliminary considerations support our modeling choices discussed in Section~\ref{sec:numerical-simul-venting}.

%%%%%%%%%%%%%%%%%%%%%%%%%%%%
%%%%%%%%%%%%%%%%%%%%%%%%%%%%
%%%%%%%%%%%%%%%%%%%%%%%%%%%%

\section{Venting simulations}\label{sec:numerical-simul-venting}

In order to quantify the contamination of the lunar environment associated with the airlock depressurization, 
we simulate the external gas expansion from a vent up to a distance of about $100~\si{m}$ from the module.
As the gas is rarefied in the entire domain, we employ here the open-source SPARTA implementation \citep{plimpton2019direct} of the 
Direct Simulation Monte Carlo (DSMC) method \citep{bird1994molecular}.

As discussed in Section~\ref{sec:depressurization}, we simulate here a mixture composed of $30\%~\mathrm{O_2}$ and $70\%~\mathrm{N_2}$.
We consider one single point along the depressurization curve.
Specifically, with reference to Fig.~\ref{fig:mdot_vent_time}, we select a mass flow rate (expressed through a single vent, so that the total is twice as much) 
of $\dot{m} = 5.0\times 10^{-7}~\si{kg/s}$, 
corresponding to a time $t \approx 10~\si{min}$.
The gas density at the vent is $\rho_v = 5.04\times10^{-6}~\si{kg/m^3}$ and the number density is $n_v = 1.04\times10^{20}~\si{m^{-3}}$.
As mentioned, the Mach number at the vent is assumed to be $M_v=1$ and the temperature is $T_v = 250~\si{K}$ at any time during the depressurization 
process, resulting in a constant bulk velocity, $u_v = 315.69~\si{m/s}$.

Numerically, the low pressure makes this the simplest condition of the expansion process 
as the gas is expected to be free-molecular in most of the external domain except at the vent, 
where the Knudsen number is $\mathrm{Kn} \approx 0.1$.
While normally this regime would be easily addressable by DSMC, the large spatial extent of the domain makes the simulations computationally demanding.
In particular, 
\begin{itemize} 
  \item A small time step is necessary in order to properly sample the collision rate at the vent,
        but it is also necessary to march in time until the particles have traveled through the large domain of interest;
  \item The grid needs to be refined to resolve the local mean-free-path and the vent geometry, requiring a significant number of simulated particles.
\end{itemize}

\begin{figure}[htpb]
  \centering
  \includegraphics[width=\columnwidth]{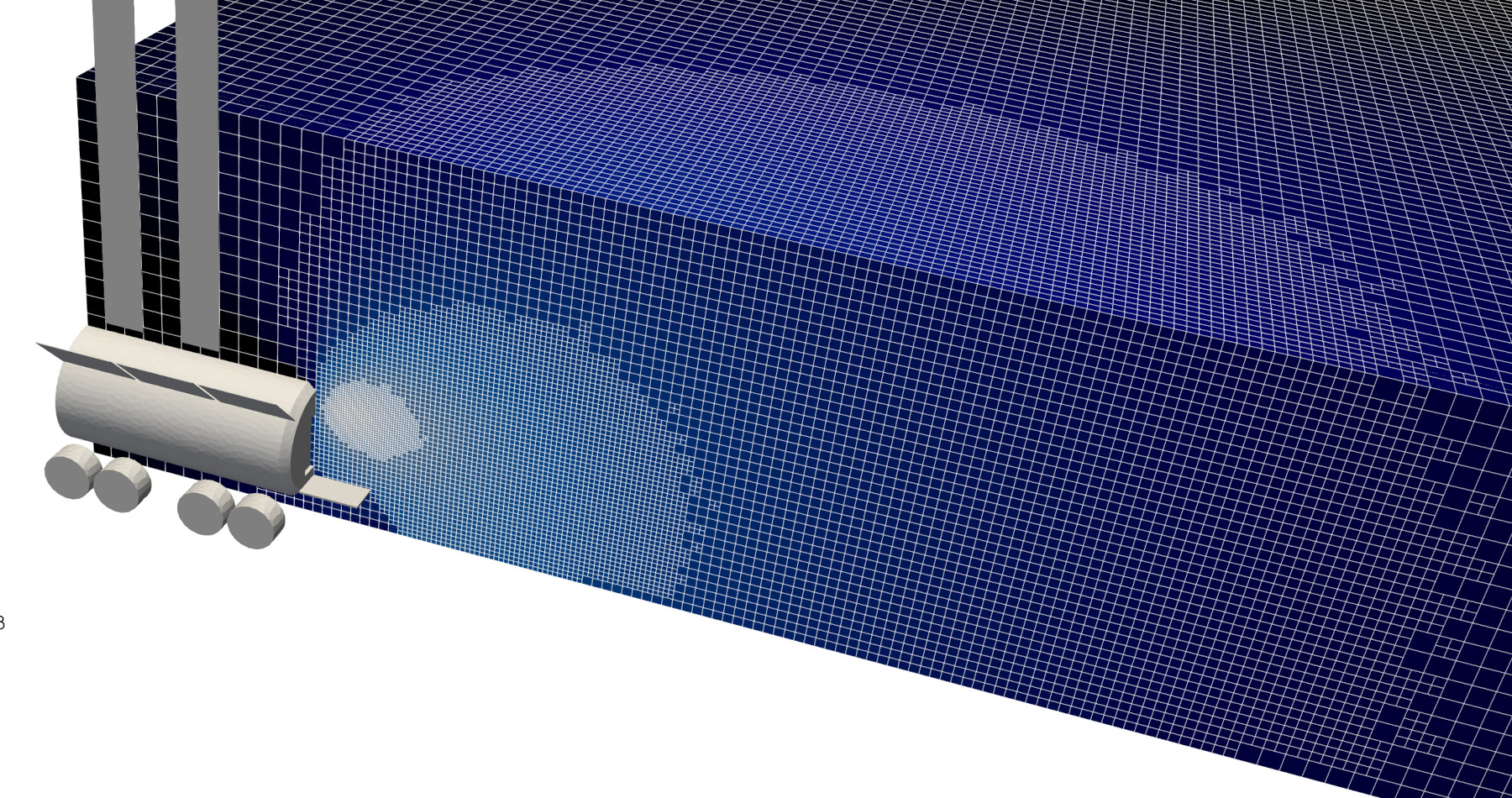}
  \caption{Detail of the computational grid employed in the front-facing vent configuration. Further refinement happens near the vent (not visible as it is embedded in the domain).}
  \label{fig:venting-grid-refinement}
\end{figure}

An example of the computational grid employed in this work is shown in Fig.~\ref{fig:venting-grid-refinement}.
In our simulations the grid is refined based on the number of simulated particles in each cell, which is proportional to the local mean free path.
Collisions between gas molecules are modeled here using the Variable-Soft-Sphere (VSS) potential \citep{bird1994molecular}, 
using the parameters pre-loaded in SPARTA and reported in Table~\ref{tab:VHS-parameters-air}.
Rotational relaxation happens with a variable relaxation number as discussed in Eq.~(A5) of \citet{bird1994molecular} evaluated with parameters also
reported in Table~\ref{tab:VHS-parameters-air}.

\begin{table}[h]
  \centering
  \begin{tabular}{r|c|c}
    & $\mathrm{N_2}$ & $\mathrm{O_2}$ \\
    \hline
    $d_\mathrm{ref}\times10^{10}~\si{[m]}$ & 4.07   & 3.96 \\
    $T_\mathrm{ref}~\si{[K]}$              & 273.15 & 273.15 \\
    $\omega$                               & 0.74   & 0.77 \\ 
    $\alpha$                               & 1.6    & 1.4  \\ 
    $Z_\mathrm{rot}^\infty$                & 18.1   & 16.5 \\
    $T^\star~\si{[K]}$                     & 91.5   & 113.5 
  \end{tabular}
  \caption{VSS and rotational-relaxation parameters employed in the venting simulations.}
  \label{tab:VHS-parameters-air}
\end{table}

Gas-surface collisions assume a diffuse scattering (full accommodation) and the temperature of every surface is set to $T_s = 300~\si{K}$ for simplicity.
The lunar regolith is simulated as a flat surface also at a uniform temperature of $300~\si{K}$. 
The actual temperature of the lunar regolith varies dramatically based on the latitude and hour of day \citep{williams2017global,boccelliUnderReviewSurfaceTemperature}. 
Additionally, shadowing caused by the body of the module, by the solar panels and the radiators could create localized low-temperature spots that might partially 
affect the local three-dimensional gas distribution.
However, besides this effect being likely small at the considered rarefaction conditions, 
it also does not affect the deposition fluxes which are only defined by the source. 
Therefore, our simplifying choice of employing a constant surface temperature does not compromise our contamination study.

The simulations are performed as follows. 
We start with no particles in the computational domain, approximating the lunar vacuum.
At each timestep particles are injected in the domain and a steady state is reached after about $0.6$ physical seconds, with
about 6 million simulated particles.
At this point, in order to reduce the statistical noise of the results, we start computing running averages of the macroscopic fields to be outputted.
Gravitational acceleration is neglected in these simulations because of the small size of the considered domain as compared to the ballistic spacial scales, 
which is in the order of tens of kilometers for air or water molecules at these temperatures.

\begin{figure*}[htpb]
  \centering
  \includegraphics[width=\textwidth]{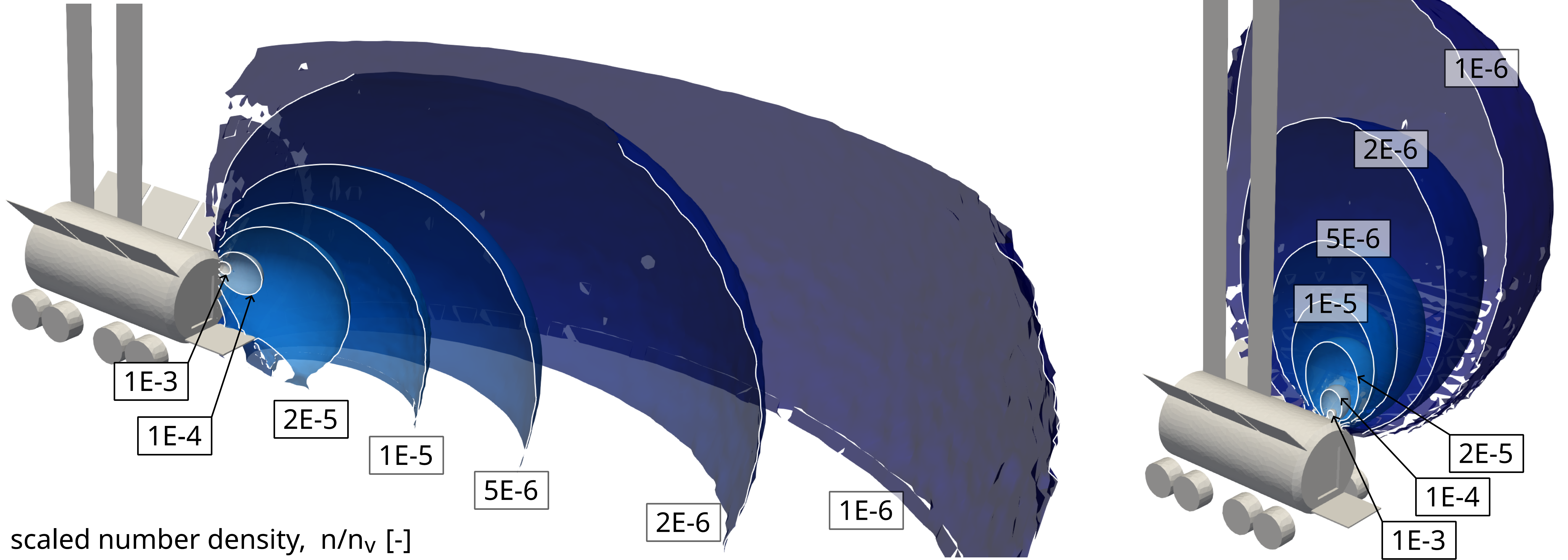}
  \caption{Contours of the number density, scaled with the vent density. Left: front-facing vent configuration. Right: top-facing vents. 
           The plots show cuts of the simulated domain through the vent center.}
  \label{fig:venting-two-config-contours}
\end{figure*}

The results of the simulations are shown in Fig.~\ref{fig:venting-two-config-contours} in terms of density contours around the vent.
Only half of the domain is shown as the simulations are symmetric.
In the figure we show the local density scaled with the density at the vent, which in the present simulations is $n_v = 1.04\times 10^{20}~\si{m^{-3}}$.

\begin{figure}[htpb]
  \centering
  \includegraphics[width=\columnwidth]{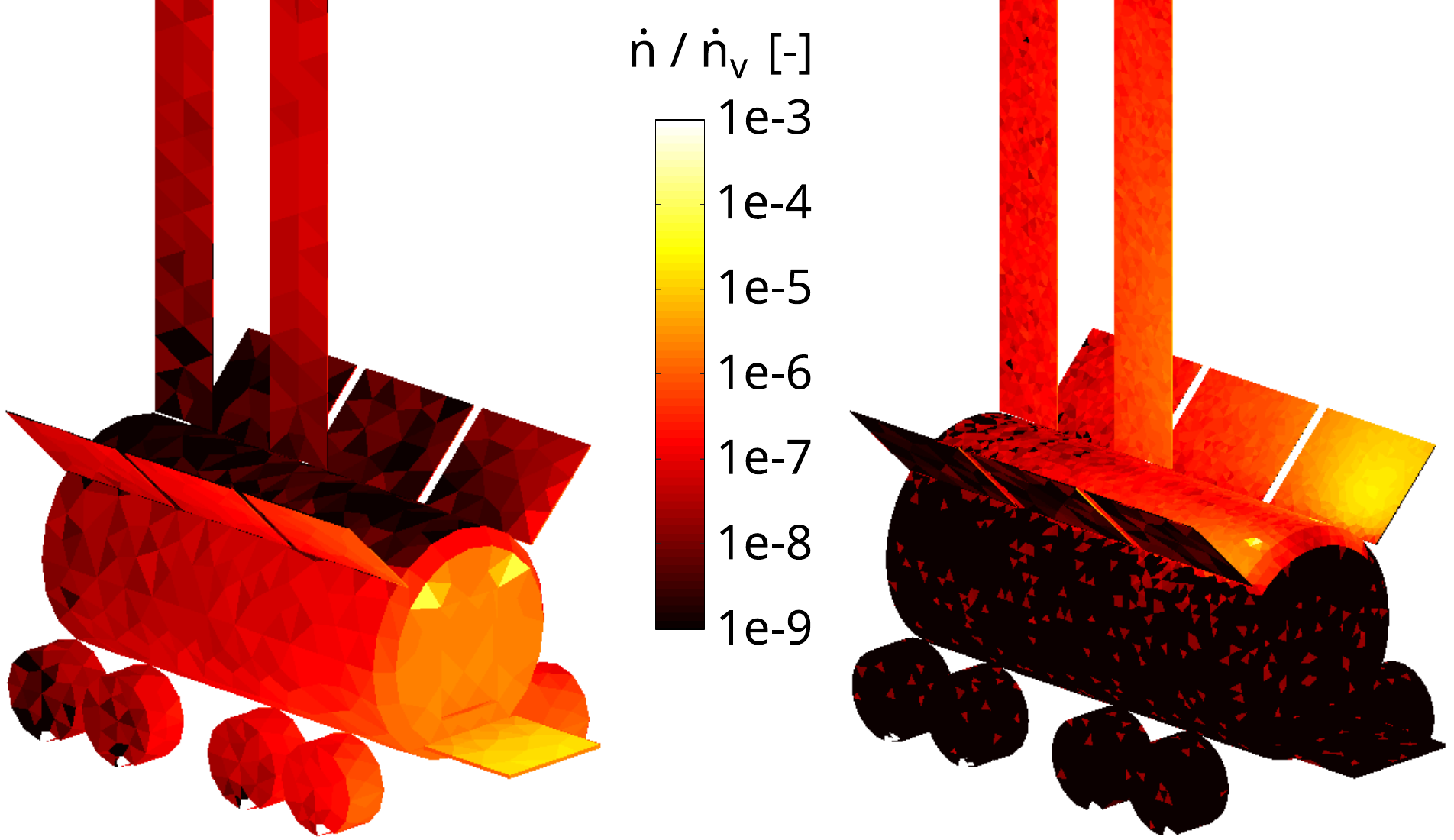}
  \caption{Incident particle flux to the module surfaces normalized with the flux injected from one vent, $\dot{n}/\dot{n}_v$, with $\dot{n}_v = \dot{m}/(m_\mathrm{mix} \pi R_v^2)$. Front-facing (Left) and  Top-facing vents (Right).}
  \label{fig:venting-two-config-surf-MPH}
\end{figure}

As the vents can be roughly interpreted as (sonic) spherical expansion points, the density is observed to drop rapidly far from them.
At a distance of $1~\si{m}$, the density has roughly dropped by four orders of magnitude and about five orders of magnitude at roughly $5~\si{m}$.
Figure~\ref{fig:venting-two-config-surf-MPH} shows the particle flux impinging on the surfaces of the module, $\dot{n}~\si{[1/(m^{2}s)]}$, 
normalized to the particle flux from the vent, $\dot{n}_v$, calculated as  $\dot{n}_v = \dot{m}/(m_\mathrm{mix} \pi R^2)$.
For the simulation considered here, $\dot{n}_v = 3.28\times10^{22}~\si{m^{-2}s^{-1}}$.
The results of Fig.~\ref{fig:venting-two-config-surf-MPH} are given in dimensionless form so that one can easily re-scale them using the desired injection flow rate, 
obtaining an estimation for the corresponding surface fluxes.

In the figure, the normalized particle fluxes are always smaller than unity, even at the vents.
This is because the plot only shows the particles \textit{impinging} on the surface, not considering the emitted ones.
At the vent location, these represent the back-scattered particles, which are roughly $10^{-4}$ times fewer than the injected ones.

\begin{figure*}[htpb]
  \centering
  \includegraphics[width=\textwidth]{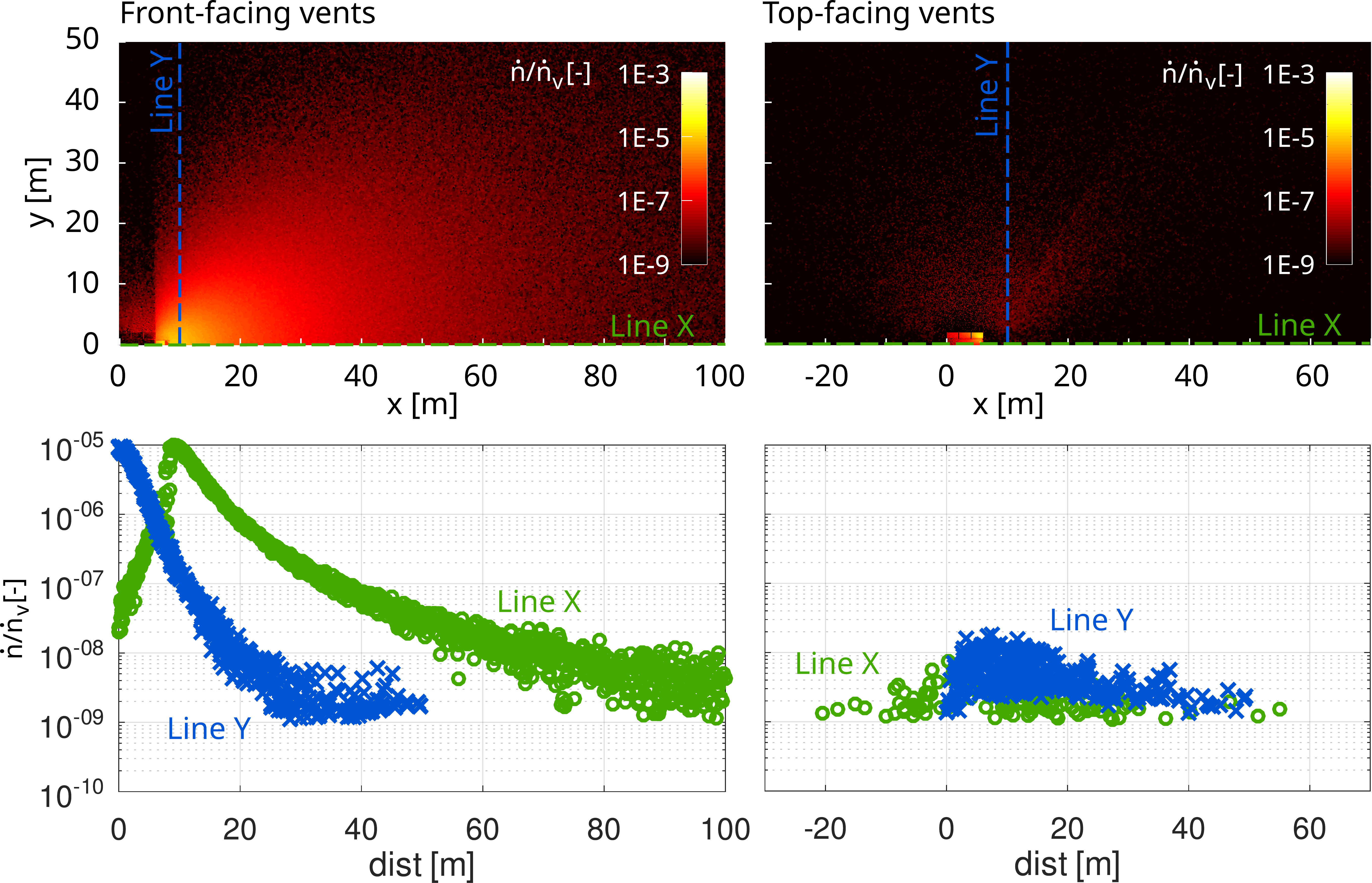}
  \caption{Top: incident particle flux at a distance from the module, normalized with the flux injected from one vent, $n/n_v$, with $\dot{n}_v = \dot{m}/(m_\mathrm{mix} \pi R_v^2)$.
           Bottom: same values extracted along the horizontal and vertical lines shown in the Top panel.
           Left boxes: front-facing vents. Right boxes: top-facing vents.}
  \label{fig:venting-two-config-surf-MPH-large}
\end{figure*}

As one might expect, the front-facing venting configuration results in a larger possible contamination of the lunar surface as compared to the top-facing one.
This is also reflected in Fig.~\ref{fig:venting-two-config-surf-MPH-large}, where we show the footprint of the contamination at a distance from the module.
To aid the interpretation of the results, the particle deposition fluxes of Fig.~\ref{fig:venting-two-config-surf-MPH-large}-Top are extracted along 
lines oriented along $x$ and $y$ and are shown in Fig.~\ref{fig:venting-two-config-surf-MPH-large}-Bottom.
The results shown in Fig.~\ref{fig:venting-two-config-surf-MPH-large} bottom out at $\dot{n}/\dot{n}_v = 10^{-9}$.
This is due to the finite number of simulated particles:
at these low values, either a single simulated particle hits the surface element, resulting in a scaled flux of $10^{-9}$, or no simulated particle hits the surface at all.
Simulations with a larger domain are possible but will require to improve the statistics by either increasing the sampling time,
by employing more simulated particles or by using a coarser surface grid.

\subsubsection*{Extrapolating the results to higher-pressure conditions}

In this section we have only considered a single pressure condition of the whole depressurization curve.
Simulating earlier conditions directly would make the computational requirements stricter:
for higher airlock pressures the vent density is higher and the Knusen number reduces as the gas approaches the continuum regime.
In such cases, one might need to employ hybrid CFD-DSMC calculations \citep{virgile2022optimisation,vasileiadis2024hybriddcfoam} and/or to
implement collision-limiter approaches \citep{titov2007extension}.
Given the preliminary nature of this work, and considering that the geometry and specific operating conditions of these modules 
are still being defined, we believe it is not necessary to reach this level of detail and we are content with our preliminary calculations.

The presented results can be scaled using the vent (injection) values such that one can extrapolate them to the 
desired pressure conditions.
The extrapolation can be done as follows:
\begin{enumerate}
  \item The desired airlock pressure, $P_a$, is selected;
  \item The associated time during the depressurization is obtained from the solid line in Fig.~\ref{fig:discharge-pressure-time};
  \item The mass flow rate, $\dot{m}$, is then found from Fig.~\ref{fig:mdot_vent_time};
  \item Recalling that the vent radius is $R_v = 0.01~\si{m}$, the molecular mass of the mixture is $m_\mathrm{mix} = 4.85\times10^{-26}~\si{kg}$ 
        and the exit velocity is $u_v=315.69~\si{m/s}$ throughout the expansion, one computes the vent number density ($n_v$) and the 
        particle flow rate through the vent ($\dot{n}_v$) as
        \begin{equation}
          \begin{cases}
            n_v       = \dot{m}/(m_\mathrm{mix} u_v \pi R_v^2) \, .\\
            \dot{n}_v = n_v u_v \, .
          \end{cases}
        \end{equation}
  \item The quantities shown in Figs.~\ref{fig:venting-two-config-contours}, \ref{fig:venting-two-config-surf-MPH} 
        and \ref{fig:venting-two-config-surf-MPH-large} can then be multiplied by $n_v$ and $\dot{n}_v$ respectively, obtaining the flow field at the desired airlock pressure.
\end{enumerate}

As stated, this extrapolation is approximated, as the flow regime changes at higher pressures, but we believe that these initial estimates provide useful quantitative insights nonetheless,
especially at this stage of the design process.

%%%%%%%%%%%%%%%%%%%%%%%%%
%%%%%%%%%%%%%%%%%%%%%%%%%
%%%%%%%%%%%%%%%%%%%%%%%%%

\section{Outgassing simulations}\label{sec:numerical-simul-outgassing}

Outgassing rates depend on 
\begin{enumerate}[i)]
  \item The operating thermal conditions of the module (bake-off of the sunlit elements and re-deposition of volatiles on cold surfaces);
  \item The specific materials and design configuration;
  \item The amount of time spent in space, which might span multiple lunar days;
  \item Airlock venting operations, which deposit fresh molecules on the (shadowed or sunlit) surfaces.
\end{enumerate}

At the present design stage the mentioned factors are not entirely defined, making an accurate outgassing prediction challenging.
Yet, we attempt here an approximated analysis based on our simplified geometrical model.

On MPH and on the Lunar Cruiser, notable sources of outgassing molecules will include the solar panels, the payloads and external structures.
The latter two will emit molecules that are eventually released 
into the lunar environment trough venting paths in the blanketing.
Additionally, the Multi-Layer Insulation (MLI) blanketing itself is a well-known outgassing source \citep{petro2014maven}.
We provide here an approximated prediction based on outgassing rates measured on other spacecraft. 

\subsubsection*{Outgassing rates from the literature}

Different spacecraft have been reported to show very different levels of outgassing (see for instance Fig.~10 in \citet{boccelli2025DSMCanalysis}).
Strict cleanliness and low-outgassing are typically necessary on scientific spacecraft, particularly if carrying sensitive instruments or if aimed at astrobiological investigations 
\citep{alred2020predicting,weiss2024operational}.
Instead, higher levels of water pressure---attributed to outgassing---have been recently detected in the exosphere surrounding commercial spacecraft 
\citep{cohen2025peregrine}.
In the case of Peregrine Mission-1, the outgassing mass flow rate, inferred from mass-spectrometric measurements, 
was estimated to be $\dot{m} = 2.85\times10^{-8}~\si{g~cm^{-2}s^{-1}}$ \citep{boccelli2025DSMCanalysis}.
This value is comparable to that of the Space Shuttle, typically attributed to the release of atmospheric humidity absorbed 
in the porous heat-shield tiles and then desorbed in space \citep{paterson1989hot,killen_temporary_2024}.

Measurements on the Midcourse Space eXperiment (MSX) satellite a few days into the mission suggested a water-molecule release rate---primarily attributed to the MLI---
of $3\times10^{-12}~\si{g~cm^{-2} s^{-1}}$ (see \citet{taylor1997early} and \citet{silver1997midcourse}).
These values are analogous to those observed on the Rosetta spacecraft \citep{schlappi2010influence}.
\citet{fugett2022contamination} also report values of roughly $10^{-12}~\si{g~cm^{-2}s^{-1}}$ for MLI under vacuum-testing conditions, 
and the main observed constituent species was water.
It should be mentioned that these rates have been shown to increse dramatically, up to 
$10^{-9}~\si{g~cm^{-2}s^{-1}}$, when the MLI was exposed to a Jupiter-like radiation environment \citep{fugett2022contamination}.
Testing of silicon-based materials \citep{soares2019high}, characteristic of the solar panels, also showed order-of-magnitude increases in the outgassing rates
under the effect of strong radiation, starting from $10^{-8}~\si{g~\cm^{-2}s^{-1}}$ during irradiation and decaying to about 
$10^{-10}$ or $10^{-11}~\si{g~\cm^{-2}s^{-1}}$ about 100 hours afterwards.
A variety of outgassed species was detected from these materials, including organic compounds in the irradiated case.
Analysis by \citet{petro2014maven} for the MAVEN spacecraft suggests a solar-panel outgassing 
rate larger than $4.6\times10^{-13}~\si{g~\cm^{-2}s^{-1}}$ and with an upper bound of $10^{-10}~\si{g~\cm^{-2}s^{-1}}$, which is roughly in line with the experiments of \citet{soares2019high}.
Measurements performed on flown and unflown solar cells of the MIR space station \citep{harvey2000outgassing} also suggested levels of outgassing of $10^{-10}~\si{g~\cm^{-2}s^{-1}}$,
and the detections suggested the presence of about 25 different species including silicone oils and hydrocarbons.
Finally, for an analysis of various other materials employed on the International Space Station the reader is referred to \citet{huang2016materials}.

\subsubsection*{Numerical simulations}
In view of the mentioned literature references, we simulate here the following outgassing rates:
\begin{itemize}
  \item The solar panels are assumed to outgas at $\dot{m} = 10^{-10}~\si{g~\cm^{-2}s^{-1}}$;
  \item The cylindrical body of the module (habitat and airlock) is assumed to be entirely blanketed with MLI and outgassing at a rate of either 
        $\dot{m} = 3\times10^{-12}~\si{g~\cm^{-2}s^{-1}}$ (representative of MSX/Rosetta measurements) or at 
        $\dot{m} =  3\times10^{-8}~\si{g~\cm^{-2}s^{-1}}$ (PITMS' measurements on Peregrine Mission-1). 
        Actual outgassing values for the future pressurized rovers are likely to be somewhere in the middle;
  \item For simplicity, the wheels, radiators and any other structures do not outgass and act as absorbing surfaces.
\end{itemize}

As outgassing results in very low densities, we compute the deposition particle fluxes using a collisionless view-factor model.
The surfaces of the module and the regolith are discretized in triangular facets and the mass flow rates reaching each surface element 
is computed geometrically assuming a cosine-law emission.

\begin{figure*}[htpb]
  \centering
  \includegraphics[width=\textwidth]{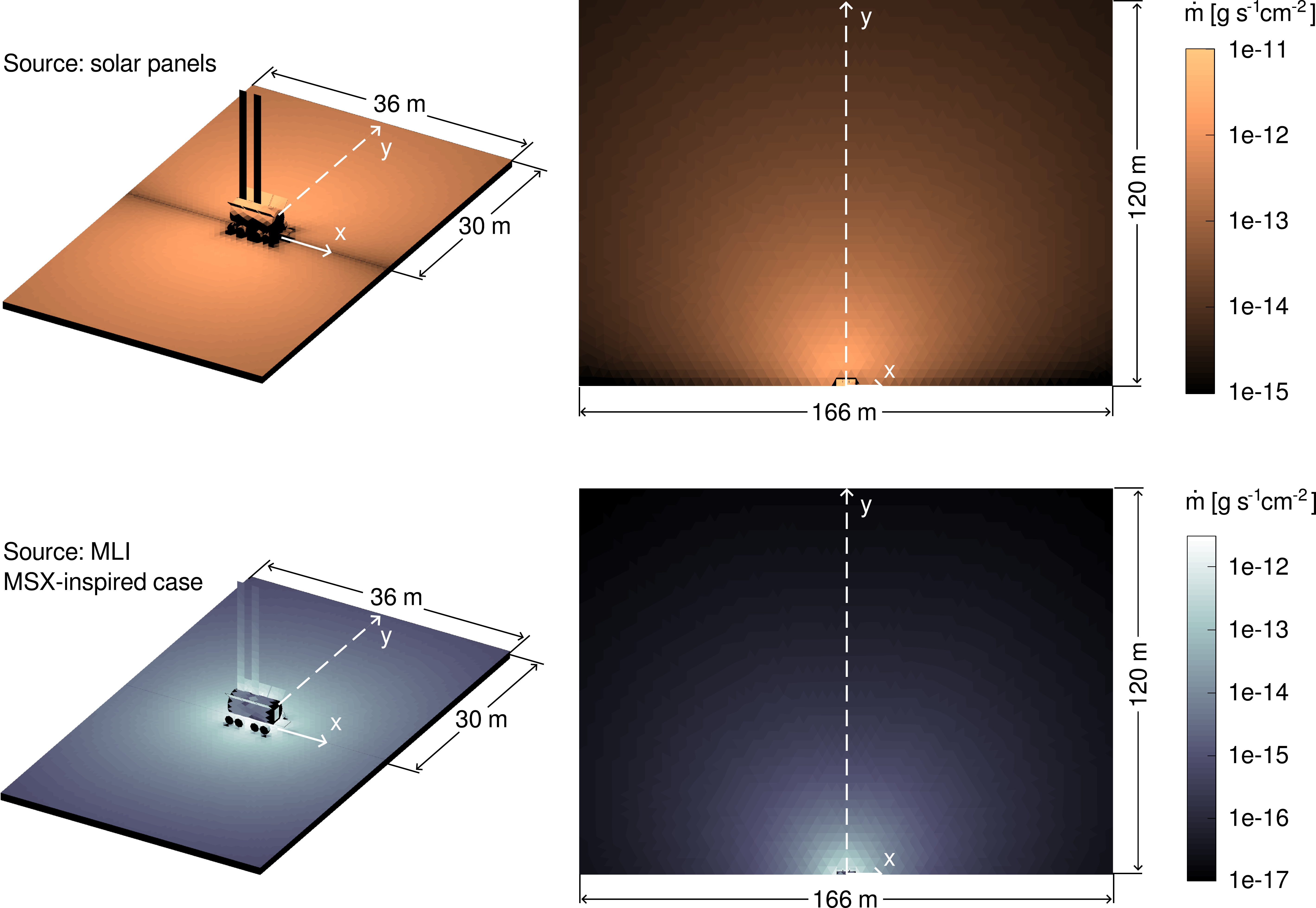}
  \caption{Outgassing simulations. Deposition rates in the vicinity of the module (Left boxes) and at a distance (Right). 
           Top: contamination from the solar panels, emitting at $\dot{m} = 10^{-10}~\si{g\,cm^{-2}\,s^{-1}}$. 
           Bottom: contamination from the MLI for an MSX/Rosetta-like emission rate of $\dot{m} = 3\times10^{-12}~\si{g\,cm^{-2}\,s^{-1}}$. 
           The case of the larger MLI outgassing rate is analogous, except that the results are four orders of magnitude larger.
           The deposition rates extracted along the white dashed line are shown in Fig.~\ref{fig:outgassing-solar-MLI-lineplot}.}
  \label{fig:outgassing-solar-MLI}
\end{figure*}

Our calculations distinguish the flux from the MLI (mainly water) from that of the solar panels (mixture of various species including organic compounds).
These different contributions are shown in Fig.~\ref{fig:outgassing-solar-MLI}.
The contamination from the MLI is mostly localized near the main body and peaks right underneath it.
Instead, contamination from the solar panels is maximum at a distance of roughly $10~\si{m}$.
This result is entirely geometrical and is due to the vertical elevation of the panels and to the shadowing 
caused by the radiator and by the main body. 

\begin{figure}[htpb]
  \centering
  \includegraphics[width=\columnwidth]{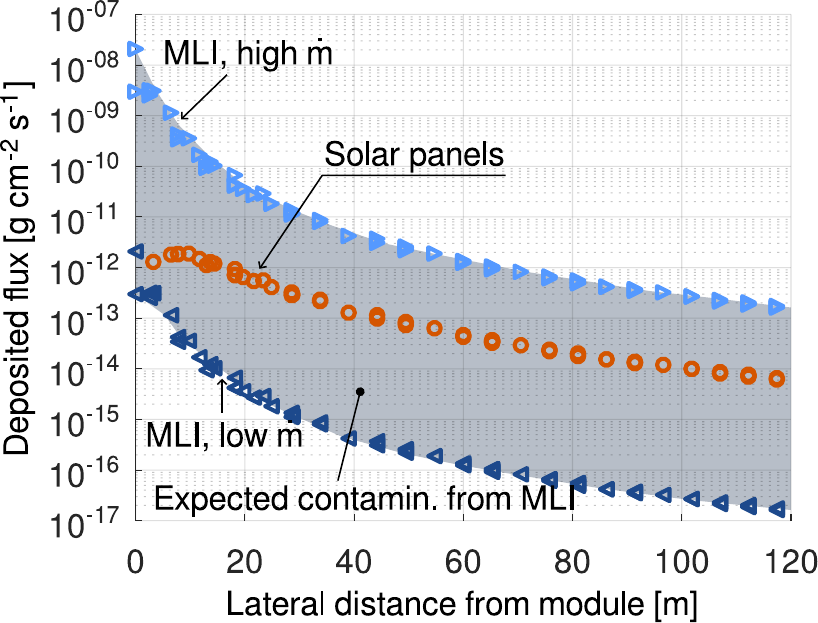}
  \caption{Lateral contamination on the lunar surface from the solar panels and the MLI (high- and low-emission scenarios).
           The actual contamination from MLI outgassing is expected to lie between the two simulated extremes (shaded gray region).}
  \label{fig:outgassing-solar-MLI-lineplot}
\end{figure}

In Fig.~\ref{fig:outgassing-solar-MLI-lineplot} we report the deposition fluxes as a function of the lateral distance from the module,
extracted along the white dashed lines of Fig.~\ref{fig:outgassing-solar-MLI}.
In this plot we include both the case of a low MLI outgassing rate and also the higher one.
The position $y=0$ represents regolith points located immediately underneath the module body.
At this location, the resulting deposition rates from the MLI can be seen to approach the values 
$2\times10^{-12}$ and $2\times10^{-8}~\si{g\,cm^{-2}\,s^{-1}}$ for the two simulations respectively.
These values coincide with the emission rate, due to the close proximity of the lunar surface to the blanketing---which constitutes 
an indirect verification of our numerical implementation.

An even larger simulation domain is analyzed in Section~\ref{sec:lunar-environment}.
However, it should be noted that the present analysis methodology cannot be spatially extended indefinitely.
At much larger distances than those considered here (say, tens of kilometers) the particle trajectory is not rectilinear anymore 
and additional fallout can be expected due to gravity.
Additionally, at such distances one might have to include other phenomena in the analysis, 
such as the ionization of outgassed molecules caused by solar radiation and the interaction with the solar wind plasma.
Such a higher-accuracy analysis goes beyond the scope of the present work but might be necessary in the case of very strict contamination requirements, 
where instruments would have to be placed at large distances from man-made structures.

%%%%%%%%%%%%%%%%%%%%%%%%%%%%
%%%%%%%%%%%%%%%%%%%%%%%%%%%%
%%%%%%%%%%%%%%%%%%%%%%%%%%%%

\section{Natural densities in the lunar exosphere}\label{sec:lunar-environment}

We can compare the numerical contamination profiles in Fig.~\ref{fig:outgassing-solar-MLI-lineplot}
with the mass flux associated with natural species in the lunar exosphere. 
Table~\ref{tab:lunar-species-fluxes} lists the surface-incident mass flux from some of the key exospheric species. 
For ${}^{40}\mathrm{Ar}$, $\mathrm{He}$, and $\mathrm{Ne}$, density values from \citet{farrell2023dust} are used
and assume an exosphere temperature of $300~\si{K}$. 
The mass flux of the water exosphere assumes a $<3~\si{cm^{-3}}$ density upper limit derived from LADEE NMS measurements \citep{hodges2022arid}.
Regarding $\mathrm{H_2}$, \citet{tucker2021effect} found molecular density values consistent with about $20\%$ 
of the incident solar wind converting to $\mathrm{H_2}$ densities consistent with \citet{hurley2017contributions}.
The solar wind influx is approximately $\approx 2\times 10^{12}~\si{m^{-2}s^{-1}}$ at the subsolar point.
In the polar regions, this value is reduced by the cosine of the solar zenith angle.
At a solar zenith angle of $80^\circ$ and assuming a $20\%$ conversion efficiency, 
the local $\mathrm{H_2}$ exospheric surface mass flux would then be $2.3\times10^{-17}~\si{g/cm^{2} s}$.

Lunar modules might deploy at a polar region situated near a permanently shadowed crater. 
Therefore, in Table~\ref{tab:lunar-species-fluxes} we also include the incident mass flux from water molecules ejected from the floor of polar craters 
onto nearby topside surfaces (i.e., ``Polar crater water''). 
This polar crater water is created by impact vaporization and solar wind sputtering of 
a $1\%$ icy regolith residing on the floor of a nearby polar crater (see Figure 4 in \citet{farrell2015spillage}).
Investigators may want to place mass spectrometers or water collecting witness plates at the edge of polar craters to analyze for this floor-released water. 

\begin{figure}[htpb]
  \centering
  \includegraphics[width=\columnwidth]{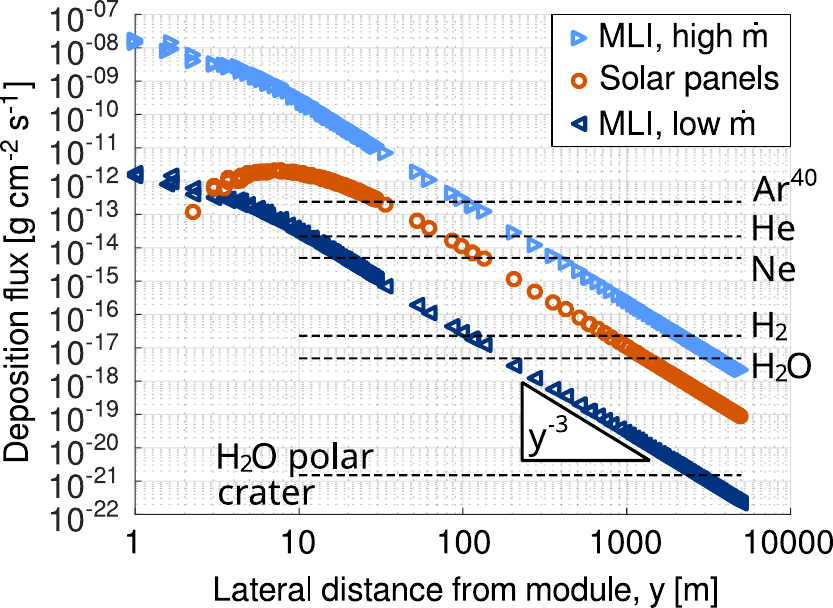}
  \caption{Deposited fluxes from the solar panels and the MLI up to $5~\si{km}$ from the surface module. Log scale. 
           The triangle indicates the expected theoretical slope $\propto y^{-3}$.
           The expected natural abundance of lunar exospheric species (Table~\ref{tab:lunar-species-fluxes}) is indicated with dashed lines.}
  \label{fig:outgassing-solar-MLI-large-lineplot}
\end{figure}

Most of the values of Table~\ref{tab:lunar-species-fluxes} are well below the contamination levels predicted in Fig.~\ref{fig:outgassing-solar-MLI-lineplot}.
Therefore, we have computed an additional view-factor simulation with a laterally-extended domain reaching a distance of $5\,000~\si{m}$.
The results are shown in Fig.~\ref{fig:outgassing-solar-MLI-large-lineplot}, where we also superimpose the expected mass flux of natural species.
At large lateral distances, $y$, the deposited fluxes are expected to decay at a rate of $y^{-3}$.
This is due to the combined effect of (i) the spherical decay (factor $y^{-2}$) and (ii) the 
sine of the relative angle between the emitting body and the receiving surface facet. 
At large values of $y$, where the body is low on the horizon, the sine can be approximated with a factor $y^{-1}$, ultimately giving the mentioned $y^{-3}$ dependence.

Considering the worst-case scenario of high-outgassing from the MLI, the contamination levels might exceed the natural ${}^{40}\mathrm{Ar}$ and $\mathrm{He}$ 
exospheric levels within about $100$ meters of the module.
In such a scenario, a simple pressure gauge sensor alone, placed within that distance, would not be able to differentiate between the module's 
contamination and the natural exosphere during the warm daytime. 
Possible solutions would be either venturing farther away from the lunar module or employing mass spectrometers.
In particular, a high-resolution mass spectrometer might be able to differentiate species via mass analysis, with the natural $\mathrm{Ar}$, $\mathrm{He}$,
and $\mathrm{Ne}$ exospheric species appearing as minor enhancements in associated mass channels relative to large contamination peaks at the
channel associated with water and hydrocarbons.
Employing species-specific high-accuracy sensors, able to measure species concentrations individually, are also a potential solution \citep{sultana2024alaska}.

For investigations targeting water molecules, operations would have to be carried further away from the lunar module.
Within $120~\si{m}$, the water exosphere would be ``swamped'' by the molecules emitted by the solar panels and by the MLI, 
even in the optimistic low-outgassing scenario. 
This distance is much larger in the case of polar crater water: 
investigators who may want to detect it may have to place instruments at a lateral distance of at least $3\,000~\si{m}$ to reduce the effect of MLI contamination.
Yet, even at such distances, the emission from solar panels or by abnormally-emitting MLI blanketing might remain prevalent.

\begin{table}[h]
  \centering
  \begin{tabular}{l|r}
    Species & Mass flux $[\si{g/cm^2\,s]}$ \\
    \hline
    ${}^{40}\mathrm{Ar}$ & $2.4\times10^{-13}$ \\
    $\mathrm{He}$      & $2.2\times10^{-14}$ \\
    $\mathrm{Ne}$      & $5.0\times10^{-15}$ \\
    $\mathrm{H_2}$     & $2.3\times10^{-17}$ \\
    $\mathrm{H_2O}$    & $<4.8\times10^{-18}$ \\
    Polar crater water & $1.5\times10^{-21}$ \\
  \end{tabular}
  \caption{Mass flux of natural exospheric species.}
  \label{tab:lunar-species-fluxes}
\end{table}

%%%%%%%%%%%%%%%%%%%%%%%%%%%%
%%%%%%%%%%%%%%%%%%%%%%%%%%%%
%%%%%%%%%%%%%%%%%%%%%%%%%%%%

\section{Conclusions}\label{sec:conclusions}

In this work we have quantified the potential contamination of the lunar surface caused by airlock venting operations and from solar-panel and Multi-Layer Insulation (MLI) 
outgassing of perspective pressurized lunar modules.

After building a representative geometry, inspired from proposed mobile lunar habitat designs, we have first analyzed the depressurization process with a simplified zero-dimensional model that represents the whole airlock as a lumped volume and assumes an isentropic
gas expansion through the valves.
We take this simplified result as an initial condition and perform a three-dimensional Direct Simulation Monte Carlo (DSMC) analysis of the further expansion of the gas into 
the lunar environment.
This is done for two different venting geometries, either facing horizontally or vertically.
The resulting self-contamination and the deposition of molecules on the lunar surface are quantified up to a distance from the module.
Horizontal vents produce a particle flux on the regolith that is about 1000 times larger than the vertical-vent configuration.

Outgassing from the solar panels and from the MLI blanketing is then studied using a view-factor model.
Representative values of the outgassing rates from the solar panels are taken from available literaure references.
For the MLI, we consider instead two limiting values: 
a low-emission scenario, for which we adopt the outgassing values reported in the literature for the MSX and for the Rosetta spacecraft, and a high-emission
scenario, for which we employ the recent measurements performed by the PITMS mass spectrometer during the Peregrine Mission-1 (similar to those estimated for the Space Shuttle).
We expect actual outgassing values for the final design of lunar modules to lie somewhere in the middle.
Two-dimensional and lateral plots of the contamination are discussed.

The analysis is performed up to a distance of $5\,000~\si{m}$ from the module and the fluxes associated with contamination are compared to the expected abundances of 
exospheric lunar species.
Scientific investigations that involve contamination-sensitive measurements should be done at a considerable distance from the module 
unless special shielding design precautions are taken.
Our preliminary analysis indicate that contamination might exceed the local abundance of argon atoms within a distance of $30$--$100$ meters from the module.
For other species, investigations would have to be carried farther away, 
and astronauts might have to venture up to and beyond $3~\si{km}$ to detect low-abundance species such as polar-crater water.

%%%%%%%%%%%%%%%%%%%%%%%%%%%%%%%%
%%%%%%%%%%%%%%%%%%%%%%%%%%%%%%%%
%%%%%%%%%%%%%%%%%%%%%%%%%%%%%%%%

\section{Acknowledgements}

The work of SB was supported by an appointment to the NPP NASA Postdoctoral Program held at NASA Goddard Space Flight Center and administered by Oak Ridge Associated Universities
under contract with NASA.
WMF's work was supported by the SSERVI CLEVER grant at Georgia Tech (NNH22ZDA020C).
OJT's work was supported by the SSERVI DREAM2 and LEADER grants.

%%%%%%%%%%%%%%%%%%%%%%%%%%%%%%%%
%%%%%%%%%%%%%%%%%%%%%%%%%%%%%%%%

% \bibliographystyle{unsrt}
\bibliographystyle{unsrtnat}
\bibliography{biblio}

\end{document}